\documentclass{article}[15pt]

\usepackage{amsmath,epsfig,color,calc,rotating,graphics}

\usepackage{natbib}
\bibpunct{(}{)}{;}{a}{}{,}

\begin{document}

\title{
Revisiting the Neutral Dynamics Derived Limiting Guanine-Cytosine Content
Using the Human De Novo Point Mutation Data
\vspace{0.2in}
\author{
Wentian Li$^1$, Yannis Almirantis$^2$,  Astero Provata$^3$\\
{\small \sl  1. The Robert S. Boas Center for Genomics and Human Genetics}\\
{\small \sl  The Feinstein Institutes for Medical Research,  Northwell Health, Manhasset, NY, USA}\\
{\small \sl  2. Theoretical Biology and Computational Genomics Laboratory, Institute of Bioscience and Applications}\\
{\small \sl  National Center for Scientific Research ``Demokritos", 15341 Athens, Greece }\\
{\small \sl 3. Statistical Mechanics and Dynamical Systems Laboratory, Institute of Nanoscience and Nanotechnology }\\
{\small \sl National Center for Scientific Research, ``Demokritos", 15341 Athens, Greece}\\
}
\date{}
}
\maketitle  
\markboth{\sl Li et al. }{\sl Li et al. }

\large

\begin{center}
Abstract
\end{center}

We revisit the topic of human genome guanine-cytosine content under neutral
evolution. For this study,  the {\sl de novo} mutation
data within human is used to estimate mutational rate instead of
using base substitution data between related species. We then 
define a new measure of mutation bias which separate the
{\sl de novo} mutation counts from the background guanine-cytosine content itself,
making comparison between different datasets easier.
We derive a new formula for calculating limiting guanine-cytosine content
by separating CpG-involved mutational events as an independent variable.
Using the formula when CpG-involved mutations are considered, 
the guanine-cytosine content drops less severely in the limit of neutral dynamics. 
We provide evidence, under certain assumptions, that
an isochore-like structure might remain as a limiting configuration
of the neutral mutational dynamics.

\vspace{0.2in}

{\bf keywords: de novo mutation, human genome, neutral evolution, 
limiting G+C content, CpG dinucleotide} 

\newpage

\section*{Introduction}

\indent

It is well known that in genomes of many species from fish \citep{costantini07} to 
human \citep{bernardi89,costantini06}, there is an alternation of DNA segments 
with high and low guanine-plus-cytosine (G+C) content, called isochores \citep{bernardi85,bernardi-book}.
Such change between high and low G+C content leads to a higher variance
with respect to DNA walk properties  than
expected from a simple homogeneous stochastic model \citep{fickett} -- thus the term
heterogeneity has been adopted \citep{sueoka,wli-yeast}. The physical spatial arrangement of
G+C content in genomes can be more complicated. Commonly observed examples of the spatial complexity
are the variable length of these isochores, the isochore-within-isochore phenomenon, and a slower
than exponential decay of base-to-base autocorrelation function, called
long-range correlation \citep{wli-kuni,ckpeng}.
An isochore is not simply a statistical signal: it has functional information also.
The richest isochores tend to be gene rich \citep{z86}, and genes tend to
have higher G+C content \citep{clay96}. G+C content may
also be associated with inner/outer region of a chromatin structure \citep{jabbari}, 
though the cause-effect relationship between the two is not completely clear \citep{wli-els}.
G+C content of  genomes of more than one thousand species is presently known \citep{wang}.

Based on the synonymous substitution rates in protein coding regions obtained
from different mammalian species, it was concluded that because G/C $\rightarrow$ A/T
substitution rate (from ancestral outgroup to descendant ingroup) 
is higher than the A/T $\rightarrow$ G/C rate, 
all genomic regions will move towards a lower equilibrium G+C content \citep{duret,belle}, though 
counter evidence was also presented \citep{alvarez,guli,romiguier}.
These data from different diverging species contain information
from a very long time scale, and mix various factors from mutation rates to selection. 

Another observation is that the G/C $\rightarrow$ A/T to A/T $\rightarrow$ G/C 
substitution rate ratio is a function of the G+C content, based on the substitutions in
DNA transposons, a type of repetitive sequences \citep{lander}. This result is
not inconsistent with \citep{duret}, where the absolute counts of
G/C $\rightarrow$ A/T substitutions over the A/T $\rightarrow$ G/C substitution
counts, increases with G+C content. 
However, there is a difference between the two approaches:
the ratio in \citep{lander} is based on conditional probability (conditional on G+C content), 
whereas the ratio in \citep{duret} is unconditional. This discrepancy is not usually emphasized,
which can cause confusion when different results are compared.

If we want to disentangle different causes, and only focus on the consequence
of short-time-scale point mutations, it would be ideal to catch 
mutational events in ``real time". Fortunately, in the genetic study of
human diseases, the ``{\sl de novo}" mutation -- genetic variant absent in parents'
genome but present in offspring's genome -- has been extensively investigated
in both diseased and normal samples \citep{kong,dutch}. The data
becomes more important as there is evidence that
{\sl de novo} mutations might play a role  in many human diseases,
such  autism spectrum disorders
\citep{sebat}, schizophrenia \citep{girard}, intellectual disability \citep{vissers,hamdan},
developmental disorders \citep{dd}, and others \citep{veltman,samocha}.

There are several distinct features of {\sl de novo} mutation data compared to the
substitutions in mammalian species: (1) a mutation occurs within one species, 
the {\sl Homo sapiens}, not mutations leading to base differences between species 
(i.e. substitution); (2) a mutation occurs in the current time, and we do not 
deal with ancestral mutational events in the past, which were likely 
to be different as it was from a potentially different environment;
(3) for a normal, disease-free offspring, a {\sl de novo} mutation
is unlikely to be deleterious, thus the requirement of neutral evolution
is satisfied. We also do not need to assume  the spread as well as fixation of 
a mutational allele when substitutions between species are used; 
(4) a mutation is more directly observed,
by comparing parents' and offspring's genome, not by comparing the inferred ancestral
genome with the descendant genomes; (5) due to availability of the human reference genome, 
not only a whole genome data can be readily obtained, but also the genomic context 
of a mutation location is well annotated, unlike the situation of
many other mammalian genomes.

The large number of {\sl de novo} studies in the human genome  in the recent years 
\citep{pranck,jonsson,goldmann,wang19,kessler}
may make the data collection daunting. Fortunately, there exist {\sl de novo} databases
which we can use directly. Note that, unlike most people who are more interested
in functional {\sl de novo} mutations that cause genetic diseases, here we are interested
in the neutral {\sl de novo} mutations in the general population. 

The study  is organized as follows: we first focus on the theoretical framework
relating {\sl de novo} mutation rate and G+C content. We re-derive the formula 
of the limiting G+C content in single-base-mutation-driven neutral dynamics in terms
of mutation counts between strong and weak  bases. 
We define coefficients $\alpha_1, \alpha_2$ 
based on mutation counts (not on mutation rate), and their deviation from the value
of 1/2 immediately reveals which type of base will increase/decrease in the limiting
dynamics.
We then expand our formula to dynamics of three variables (weak bases, strong bases
not-involved in CpG dinucleotides, and strong bases involved in CpG dinucleotide). 
We define coefficients $\beta_1, \beta_2, \beta_3$ based on mutation counts among these
three variables, and a deviation of  their  value from 1/3 immediately predicts
the limiting dynamics of them. 
We carry out a data analysis using the {\sl de novo} mutation database to
estimate $\alpha$'s and $\beta$'s in different location types, as well as in
different G+C backgrounds.

\section*{Theoretical formulation}

\subsection*{Formula of limiting G+C content based on mutation-driven neutral dynamics}

\noindent

Assume that a certain number (= $n_{WS}$) of weak (A or T) to strong (C or G) {\sl de novo} mutation events 
is observed, and similarly there are  $n_{SW}$ strong-to-weak mutation events. 
Denote the G+C-content as $x$; assume that  there are $N$ genomic positions
to be considered, and $M$ is the number of persons from which {\sl de novo}
mutation events were collected, then there are $M \cdot N \cdot x$ positions occupied by strong bases,
and $M  \cdot N \cdot (1-x)$ positions by weak bases.  We then compose the mutation (and non-mutation)
event counts matrix as:
\begin{equation}
\label{eq-ws}
\begin{array}{cc}       
&
\begin{array}{cc}
       W=A/T &  S=C/G  \\
\end{array}
\\
\begin{array}{cc}
W=A/T \\
S=C/G 
\end{array}
 &
\left(
\begin{array}{cc}
M  N (1-x) -n_{WS} & n_{WS}   \\
n_{SW} & M N  x - n_{SW}  \\
\end{array}
\right)
\end{array}
\end{equation}
The diagonal elements in Eq.(\ref{eq-ws}) are mostly  not directly counted, 
simply because (e.g.) $A \rightarrow A$ is not reported as a mutation event
(though (e.g.) $A \rightarrow T$ is). However, we can infer
them from the total number of base positions $N$, total number of samples
$M$,  the current G+C content $x$, and the mutation counts away from the
base type.

Normalizing the matrix in Eq.(\ref{eq-ws}) by row sum, we obtain the conditional 
probability (transition probability in Markov chain):
\begin{equation}
\label{eq-transition}
\begin{array}{cc}       
&
\begin{array}{cc}
       W=A/T &  S=C/G  \\
\end{array}
\\
\begin{array}{cc}
W=A/T \\
S=C/G 
\end{array}
 &
\left(
\begin{array}{cc}
 1- \frac{n_{WS}}{MN(1-x)} & p_{W \rightarrow S} \equiv \frac{n_{WS}}{M  N (1-x)}   \\
p_{S \rightarrow W} \equiv \frac{n_{SW}}{M  N x} &  1-\frac{n_{SW}}{M N x}   \\
\end{array}
\right)
\end{array}
\end{equation}
From the two  weak $\leftrightarrow$  strong (conditional) transition probabilities, it is well known
that the limiting G+C (strong) content is \citep{sueoka,petrov,lynch-book,lynch-pnas,lander,wli-jtb,wli-els}:
\begin{equation}
\label{eq-limit0}
x' = \frac{1}{ \frac{p_{S \rightarrow W}}{ p_{W \rightarrow S} }+1}.
\end{equation}
An easy derivation is to consider the ``detailed balance": $x' p_{S \rightarrow W} = (1-x') p_{W \rightarrow S}$.

We define two new coefficients based on the mutational event counts:
$\alpha_1= n_{SW}/(n_{SW}+n_{WS})$
and $\alpha_2= n_{WS}/(n_{SW}+n_{WS})$. Note $\sum_i \alpha_i=1$. 
The conditional transition probability in Eq.(\ref{eq-limit0}) can be replaced
by the {\sl de novo} mutational event counts
($n_{SW}$ and $n_{WS}$) or $\alpha$'s:
\begin{equation}
\label{eq-limit}
x' = \frac{1}{\frac{n_{SW}}{n_{WS}} \cdot \frac{1-x}{x} +1}.
= \frac{1}{\frac{\alpha_1(x)}{\alpha_2(x) } \cdot \frac{1-x}{x} +1}.
\end{equation}
Both total number of bases $N$ and number of persons $M $ are canceled 
from Eq.(\ref{eq-limit}), thus we do not need to know their values.
Eq.(\ref{eq-limit}) shows how the limiting G+C content (x') depends on the current G+C content (x),
and two (actually one) mutational count based coefficients $\alpha_1$ ($\alpha_2=1-\alpha_1$).
Eq.(\ref{eq-limit}) can be written in a more symmetric form:
\begin{equation}
\label{eq-limit2}
\frac{x'}{1-x'} =  \frac{n_{WS}}{ n_{SW}} \cdot \left( \frac{x}{1-x} \right)
= \frac{ \alpha_2(x) }{ \alpha_1(x) } \cdot \left( \frac{x}{1-x} \right).
\end{equation}
As long as $n_{WS} < n_{SW} $ (or $\alpha_2 < \alpha_1$, more C/G to C/T mutation events than in the opposite direction),
$x/(1-x)$ will decrease, so will G+C-content x. 

Note that Eq.(\ref{eq-limit2}) is not a one-time iteration, from time $t$ to time $t+1$,
typically seen in the field of dynamical systems \citep{may,li-yorke,feigenbaum}.
Eq.(\ref{eq-limit2}) maps directly from the current G+C-content to the limiting G+C-content in one step.
The base transition counts ($n_{SW}$ and $n_{WS}$) or their normalized values ($\alpha_1$ and $\alpha_2$)
are not constant values, but changing as  a function of the current G+C-content. If G+C-content reduces, 
we should also see a lower value of $\alpha_1$. To emphasize this point,  we write this
functional dependence  of $\alpha_1$ and $\alpha_2$ on $x$  in Eqs.(\ref{eq-limit},\ref{eq-limit2}) explicitly.

In the literature, the mutational bias towards W=A+T is defined as
$m= p_{S \rightarrow W}/ p_{W \rightarrow S} =v/u$ \citep{lynch-book} (page-126),
or the equilibrium constant in the direction of W base pairs $K$ \citep{lander} (page-886),
whereas their dependence on G+C-content is not obvious. In our notation,
$K= [\alpha_1(x)/\alpha_2(x)] \cdot [(1-x)/x]$ is expressed in two parts so that it is made explicit that 
the first part is derived purely from the mutation counts $\alpha_1/\alpha_1= n_{SW}/n_{WS}$
and the second part is unrelated to mutation counts, but purely base composition related.

There is another advantage of using $\alpha_{1,2}$ instead of the mutational bias $K$.
When the values of $\alpha_{1,2}$  are compared to 0.5, we immediately know the
direction of the base type change:   if $\alpha_1 > 0.5$, (A+T)-content will
increase from the current value to its limiting value; similarly if $\alpha_2 > 0.5$,
(G+C)-content will increase. This advantage is more clear in the next subsection when 
three variables are considered.

\subsection*{Formula of limiting G+C content when CpG is considered separately}

\indent

Now we specifically consider a subset of strong bases within the 5'-CpG-3' dinucleotide context.
The base C next to a base G in downstream (3') direction is known to have a much higher
mutation rate (in particularly, to base T). The dinucleotide on the opposite strand 
of 5'-CpG-3' is also  5'-CpG-3',  but the base G is expected to have a higher mutation rate. 
Let's denote these strong bases
as $S_p$ ($p$ indicates the phosphodiester bond between C and G) and other G/C bases not in this context as $S_n$.
We also assume among strong G/C bases, a proportion of $y$ of them are in $S_p$. Though 
not common, it is still possible to have a mutation from one $S_p$ base to another $S_p$ base, 
e.g., $5'-CGG-3' \rightarrow 5'-CCG-3'$.

\begin{figure}[t]
 \begin{center}
   \epsfig{file=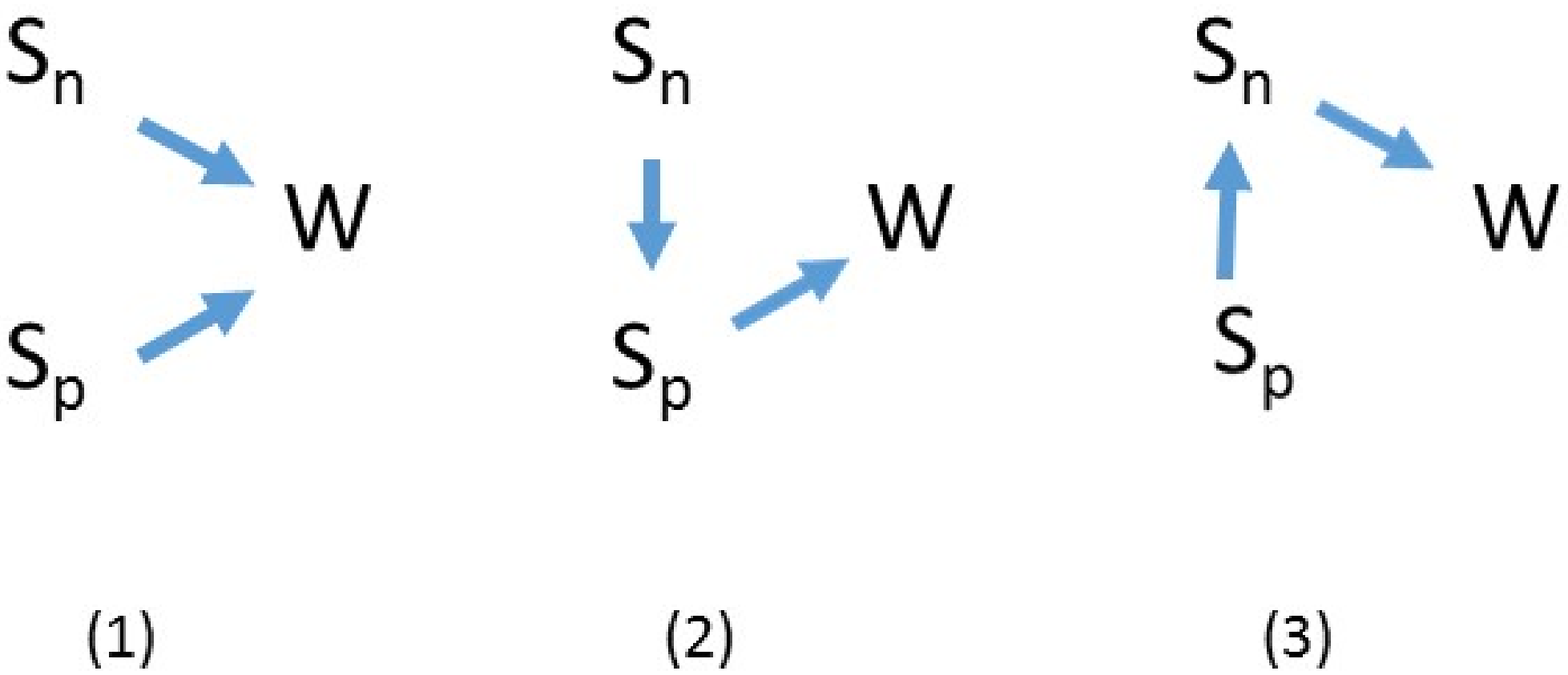, width=13cm}
 \end{center}
\caption{ \label{fig1}
Illustration of the analytic expression of the first element in the limiting array
in Eq.(\ref{eq-limit3}), which is proportional to $\beta_1$. This expression is the sum
of three terms:  $n_{SpW} n_{SW}$, $n_{SnSp} n_{SpW} $, and  $n_{SpSn} n_{SnW}$ 
where $W$ is weak base (A or T), $S_n$ is strong base (C or G)
not involved in a CpG context, and $S_p$ for S in a CpG context. These three terms
can be represented by the subplots (1), (2) and (3) .
}
\end{figure}

Similar to Eq.(\ref{eq-ws}), the number of mutation counts for three types of base (W, $S_n$, $S_p$)
are:
\begin{equation}
\label{eq-ws3}
\begin{array}{ccc}       
&
\begin{array}{ccc}
       W \hspace{1in}  & \hspace{0.5in} S_n \hspace{0.5in}  & \hspace{1in}  S_p  \\
\end{array}
\\
\begin{array}{ccc}
W \\
S_n  \\
S_p
\end{array}
 &
\left(
\begin{array}{ccc}
M  N (1-x) -n_{WSn}-n_{WSp} & n_{WSn} & n_{WSp}   \\
n_{SnW} & M N x (1-y) - n_{SnW} -n_{SnSp} & n_{SnSp}  \\
n_{SpW} & n_{SpSn} & M N x y - n_{SpW} -n_{SpSn}  \\
\end{array}
\right)
\end{array}
\nonumber
\end{equation}
and again, the row normalized matrix is the transition matrix:
\begin{equation}
\label{eq-transition3}
\begin{array}{ccc}       
&
\begin{array}{ccc}
       W \hspace{0.5in}  & \hspace{0.5in} S_n \hspace{0.5in}  & \hspace{0.5in}  S_p  \\
\end{array}
\\
\begin{array}{ccc}
W \\
S_n  \\
S_p
\end{array}
 &
\left(
\begin{array}{ccc}
1- \frac{n_{WSn}+n_{WSp}}{MN(1-x)} & \frac{n_{WSn}}{MN(1-x)} & \frac{n_{WSp}}{MN(1-x)}   \\
\frac{ n_{SnW}}{MNx(1-y)} & 1 - \frac{n_{SnW} +n_{SnSp}}{MNx(1-y)} & \frac{n_{SnSp}}{MNx(1-y)}  \\
\frac{ n_{SpW}}{MNxy} & \frac{n_{SpSn}}{MNxy} & 1- \frac{n_{SpW} +n_{SpSn}}{MNxy}  \\
\end{array}
\right)
\end{array}
\end{equation}
The limiting composition of $W, S_n, S_p$ is proportion to the eigenvector of the
transpose (switching rows and columns) of Eq.(\ref{eq-transition3}) corresponding
to the eigenvalue equal to 1 (which is the largest eigenvalue of a Markov transition matrix)
(see Appendix). 
We obtain such  a (unnormalized) eigenvector for the transpose of  Eq.(\ref{eq-transition3})
by Wolfram Alpha ({\sl www.wolframalpha.com}) as:
\begin{equation} 
\label{eq-limit3}
\left(
\begin{array}{c}
(n_{SpW} n_{SnW} + n_{SnSp} n_{SpW} + n_{SpSn} n_{SnW} ) (1-x) \\
(n_{WSn} n_{SpSn} + n_{WSp} n_{SpSn} + n_{SpW} n_{WSn} ) x (1-y) \\
(n_{WSp} n_{SnSp} + n_{WSn} n_{SnSp} + n_{SnW} n_{WSp}) x y \\
\end{array}
\right)
\propto
\left(
\begin{array}{c}
\beta_1(x) \cdot  (1-x) \\
\beta_2(x) \cdot x (1-y) \\
\beta_3(x)  \cdot x y \\
\end{array}
\right).
\end{equation} 
Although Eq.(\ref{eq-limit3}) looks complicated, it can be memorized by the illustration in Fig.\ref{fig1}. 
Note that again the genome size N and number of persons M are not present in the limiting
composition formula Eq.(\ref{eq-limit3}). 

We introduce new coefficients $\beta_i$ (i=1,2,3) to be proportional to the coefficients in the 
left column-array in Eq.(\ref{eq-limit3}), but normalized
(i.e.,  divided by the sum of all products of two transition counts). 
Note that $\sum_i \beta_i =1$ doesn't mean the right column array itself in
Eq.(\ref{eq-limit3}) is normalized.  Our introduction of $\beta_i$ coefficients
will make the comparison between different data easier, because they are based purely on mutational
counts. Also, if $\beta_1=\beta_2=\beta_3=1/3$ implies no change in the
composition of weak, CpG-unrelated-strong, and CpG-related-strong base types,
and deviation of $\beta$ from 1/3 easily points to the direction of change in the composition.
To emphasize the fact that $\{ \beta_i\}$ are not constant in the dynamics, 
we write their dependence on G+C-content explicitly in Eq.(\ref{eq-limit3}).

\section*{Data analyses}

\subsection*{Filtering neutral {\sl de novo} mutation events}

\noindent

We use the denovo-db v1.6.1 
({\sl http://denovo-db.gs.washington.edu/denovo-db/}, August 19, 2018). The files
denovo-db.ssc-samples.variants.tsv and denovo-db.non-ssc-samples.variants.tsv are used. 
Each line in these files is a mutational event in a person with a particular annotation. 
Therefore, for a mutation in a coding region with multiple transcripts, each mutation
event may occupy multiple lines. There are 628,234 lines in the two files.
The genomic coordinates are in hg19/GRCh37.

We filter the {\sl de novo} mutations by the following criteria: 
(1) The mutation is a bi-allelic single-nucleotide-polymorphism (SNP); 
(2) The person's phenotype is a normal ``control";
(3) Y-chromosome variants are excluded; 
(4) the base is consistent with the reference genome of hg19/GRCh37.
The criterion  \#1 serves to avoid the more complicated mutational events such as 
indels and multi-allelic variants whose detection is less reliable.
The criteria  \#2 is to make sure that the mutational event is neutral, and less likely
to be deleterious. These filterings reduce the number of  lines to 110,989.

The following studies have contributed the most to the {\sl de novo} mutation event counts:
Turner2017 (83187, 75.0\%) \citep{turner17}, 
GONL (15896, 14.3\%) \citep{gonl}, 
Turner2016 (3541, 3.2\%) \citep{turner16}, 
Iossifov2014 (3521, 3.2\% ) \citep{iossifov}, 
Werling2018 (3302, 3.0\% ) \citep{werling}, 
Krumm (1014, 0.9\%) \citep{kurmm}, 
Yuen2017 (181, 0.16\%) \citep{yuen17},
Gulsuner2013 (170, 0.15\%) \citep{gulsuner}, 
Conrad2011 (59, 0.05\%) \citep{conrad},
Besenbacher2014 (52, 0.047\%) \citep{besen},
Rauch2012 (509, 0.045\%) \citep{rauch},
and ASD3 (15, 0.014\%).

Besides the information provided by denovo-db, we have added these extra information by
using the hg19/GRCh37 reference genome: (1) G+C base and CpG dinucleotide count 
of 2kb window centered at the SNP; (2) G+C base and CpG dinucleotide count  of 20kb window 
centered at the SNP; (3) the triplet context
of the SNP; and (4) the triplet context after the mutation.

The denovo-db provides 18 location-types which we condense to 9 types:
intron (and intron-near-splice): 58824 lines, intergenic: 40375 lines,
upstream-gene and downstream-gene: 4832 lines,
missense(and missense-near-splice): 3235 lines,
5' and 3' UTR: 1514 lines,
synonymous (and synonymous-nea-splice): 1336 lines,
non-coding-exon (and non-coding-exon-near-splice): 634 lines,
stop-lost and stop-gain: 149 lines,
splice-acceptor and splice-donor: 81 lines.

\begin{figure}[t]
 \begin{center}
  \begin{turn}{-90}
   \epsfig{file=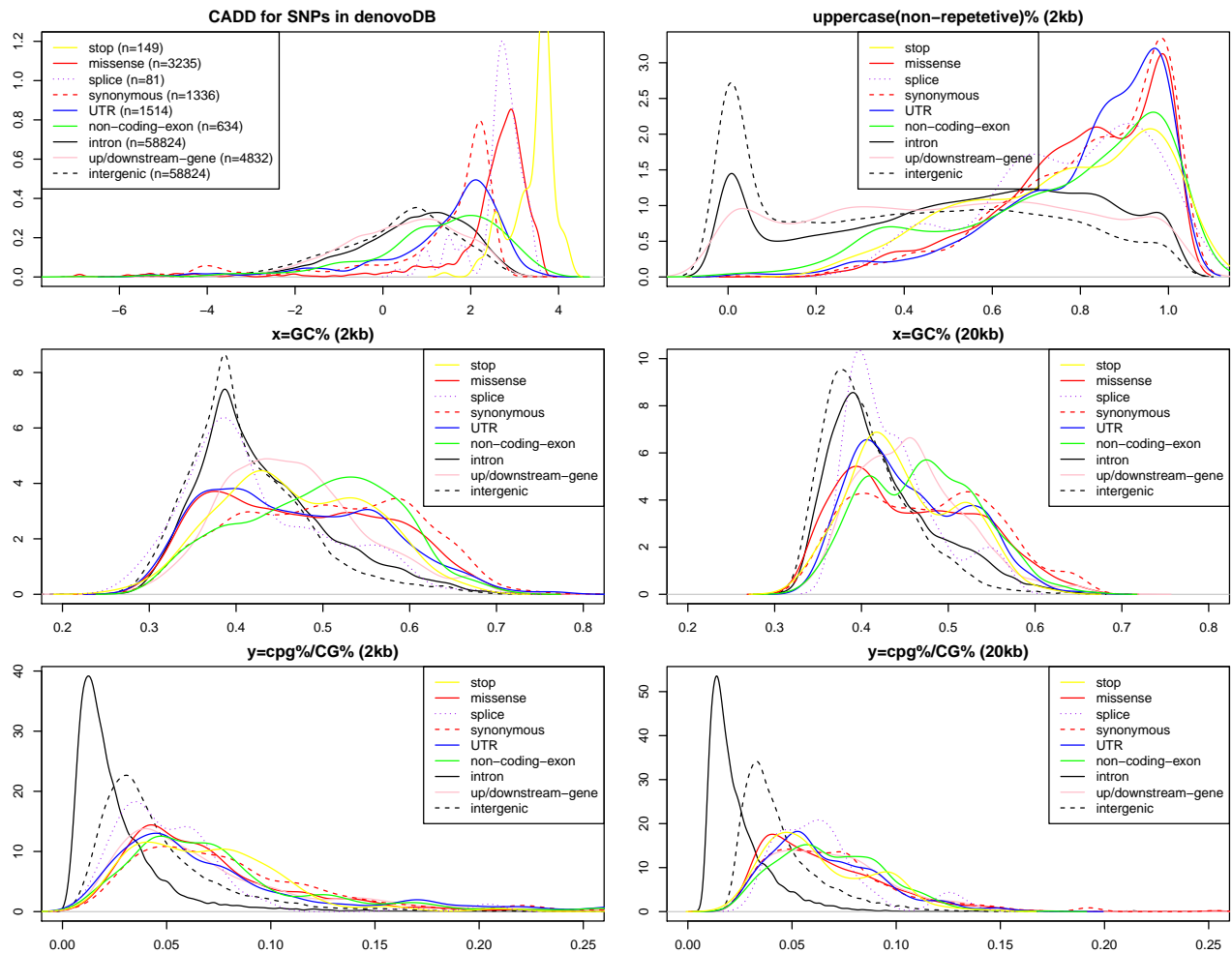, width=13cm}
  \end{turn}
 \end{center}
\caption{ \label{fig2}
Distribution of various statistics of the {\sl de novo} mutations according to nine different categories:
stop-gain/loss, missense, splice donor/acceptor, synonymous, 3'/5'-UTR, 
non-coding-exon, intron, up/downstream-gene, intergenic.
(1) CADD; (2) percentage of non-repetitive sequence in the 2kb window centered at the mutation site ;
(3) (G+C)-content in the 2kb window;
(4) (G+C)-content in the 20kb window;
(5) CpG/(G+C) in the 2kb window;
(6) CpG/(G+C) in the 20kb window.
}
\end{figure}

Fig.\ref{fig2} shows the distribution of CADD (combined annotation dependent depletion) 
value \citep{cadd}, percentage of non-repetitive-sequence (uppercase letter),
2kb window G+C content, 20kb window G+C content, 2kb widow CpG\%/(G+C)\%, and 20kb widow CpG\%/(G+C)\%
of all these 9 location types. Most of the result in Fig.\ref{fig2} is known. For example,
the functional impact of variants is the highest for stop-gain/lost, followed by missense;
intergenic regions contain more repetitive sequences or transposons; genic regions can be of
high-(G+C)-content; etc. We further show that larger window (20kb) statistics have more narrow
distributions, and intron regions (even more so than intergenic regions) avoid CpG dinucleotides.

\subsection*{De novo mutation derived $\alpha$ and $\beta$ coefficients}

\noindent

Table \ref{table1} shows the raw count of different types of {\sl de novo} mutations in 9 different
variant types described in the last section. The two $\alpha_1$ and $\alpha_2$ coefficients for the two-base type
model and three $\beta_1, \beta_2, \beta_3$ coefficients for the three-base type model are listed in Table \ref{table1}.
Although we cannot assume neutral dynamics for variants in the functional categories,
whether an $\alpha_i$ coefficient larger or smaller than 1/2, and whether a $\beta_i$ coefficient
larger or smaller than 1/3 will indicate which direction the mutational force is pushing.
In all functional categories, W (A+T) base content will be pushed higher by mutation,
S(G+C) and CpG content will be pushed lower. The stop-gain/loss and splice acceptor/donor
categories contain very few {\sl de novo} mutation counts. However, the mutational force would
drive the CpG content higher in splice sites, while deplete CpG from stop sites.

Bases in the intergenic regions can be assumed to follow a neutral evolution without constraints.
We further partition the {\sl de novo} mutational events in intergenic regions according to its
surrounding (2kb)  (G+C)-content, and $\alpha$, $\beta$ coefficients are calculated in each (G+C)-content 
quantile. The results are shown in Table \ref{table2}. We can see that not only $\alpha_1 > 1/2$ and
$\beta_1 > 1/3$, but also their values increase with the surrounding G+C content. This result is consistent
with previous publications \citep{duret,lander}. Table \ref{table2} also shows that $\beta_3 < 1/3$,
and decreases with surrounding G+C content. Dependence of CpG mutation rate on local G+C content
has also been reported before \citep{fryxell}.

\subsection*{Evidence of two different limiting G+C contents}

\indent

\begin{figure}[t]
 \begin{center}
  \begin{turn}{-90}
   \epsfig{file=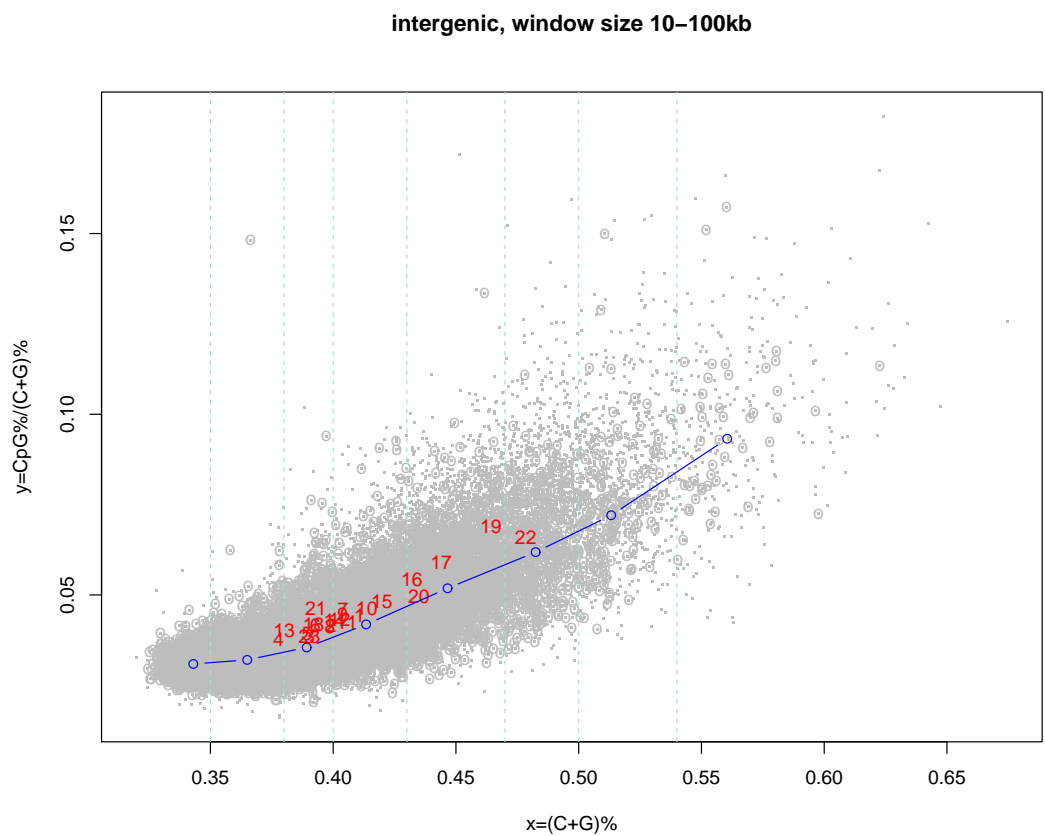, width=11cm}
  \end{turn}
 \end{center}
\caption{ \label{fig3}
Each point is an intergenic region in the human genome (regions longer than 100kb are split into 
multiple 100kb windows),
where x is G+C content, y is the CpG/(G+C) proportion. Larger windows/regions ($>$ 80kb) 
are represented by circles. The x=G+C is partitioned into 6 ranges, where the blue dots represent the bin average.
The chromosome-level averages are indicated by red letters.
}
\end{figure}

To further examine the prediction of our neutral mutational dynamics, using the 
mutation rates based on the {\sl de novo} mutational event count, as a function
of current G+C content, we expand the previous six G+C content quantiles to eight,
with the highest G+C range split into three more G+C regions. This partition would
lead to around 6000-7000 intergenic {\sl de novo} mutational events in each one of the lower G+C brackets,
but 2000-3000 intergenic mutational events in the last three high G+C brackets. The mutation counts
of various types, the calculated $\alpha(x)$'s and $\beta(x)$'s, the predicted
limiting G+C content (by either two-variable or three-variable equation)  and 
limiting CpG/(G+C), are shown in Table \ref{table2}. The current intergenic G+C content
and CpG/(G+C) values, calculated directly from the hg19 intergenic sequences 
(an intergenic sequence longer than 10kb is partitioned into pieces of 10kb length),
are shown in Fig.\ref{fig3}. We notice that CpG/(G+C) is positively correlated with
(G+C) \%, as it involves  the product of two strong bases.

\begin{figure}[t]
 \begin{center}
  \begin{turn}{-90}
   \epsfig{file=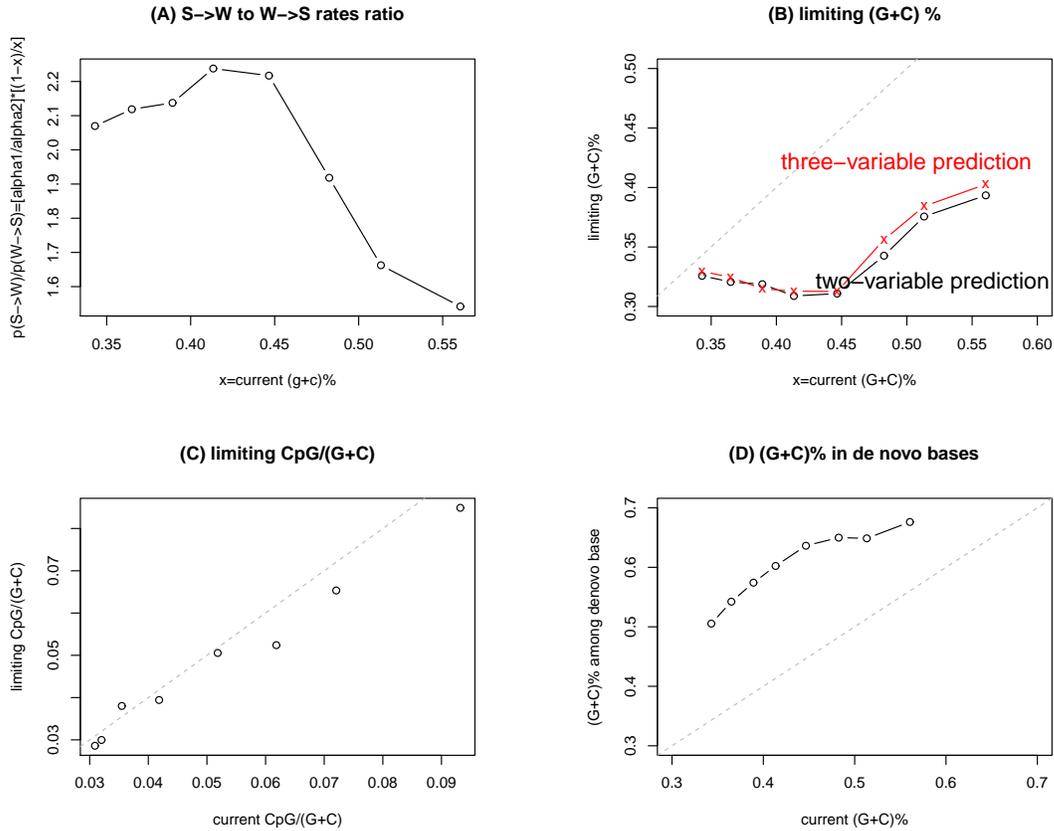, width=11cm}
  \end{turn}
 \end{center}
\caption{ \label{fig4}
Evidence that current isochore structure might be maintained in the equilibrium configuration
in neutral dynamics. Each bin represents a collection of {\sl de novo} events in intergenic
regions with specific 2kb window G+C content. The first five G+C bin points contain
6000-7000 mutational events each, whereas the last 3 high G+C bin points only 2000-3000 
mutational events each.
(A) $\frac{p_{S \rightarrow W}}{ p_{W \rightarrow S} }$ (Eq.(\ref{eq-limit0})) or
$ \frac{\alpha_1(x)}{\alpha_2(x) } \cdot \frac{1-x}{x}$ (Eq.(\ref{eq-limit}))
as a function of current 2kb G+C content;
(B) limiting G+C by Eq.(\ref{eq-limit}) or Eq.(\ref{eq-limit3}) as a function of current 2kb G+C content;
(C) limiting CpG/(G+C) by Eq.(\ref{eq-limit3}) as a function of current 2kb CpG/(G+C);
(D) proportion of G+C among bases which have experienced {\sl de novo} mutation.
The grey lines in (B), (C), (D) represent slope=1 diagonal lines.}
\end{figure}

Fig.\ref{fig4} depicts the mutational data as a function of current (G+C)\% from 
various perspectives. Fig.\ref{fig4}(A) shows that the $p(S \rightarrow W)/p(S \rightarrow W)$
is not constant, but decreases at high (G+C) content. By the prediction in Eq.(\ref{eq-limit0}),
the limiting (G+C)\% will be higher for the current (G+C)-rich intergenic regions,
as shown in Fig.\ref{fig4}(B). Interestingly, the three-variable prediction (by
considering CpG-containing strong base as a distinct variable) leads to slightly
higher limiting (G+C)\% than the two-variable prediction. Fig.\ref{fig4}(C), however,
shows that the limiting CpG/(G+C) may not be off very much from the current CpG/(G+C)
value. Fig.\ref{fig4}(D) is yet another way to look at the same data. If
the mutational rate is the same in regions with different current (G+C)\%,
the proportion of G or C bases among those that experience a {\sl de novo} mutation
will be linearly proportional to the current (G+C)\%. Nonlinearity of this
``substituted-base G+C content" in bacterial genomes has been observed \citep{bohlin}.
The fact that our Fig.\ref{fig4}(D) is further away from the diagonal line
(see the grey dashed line) indicates that human genome is not in a base composition 
equilibrium state as in bacterial genomes.

The results in Fig.\ref{fig4} may indicate that isochores are maintained by
neutral mutational dynamics if the mutational rate is estimated from the {\sl de novo} mutational 
events. However, there are still two possibilities: (1) the relatively low mutational
AT-driving-force observed in the current high G+C region is supposed to be still
low when the G+C content in the same region is lower in time. We may justify
this assumption by a hypothesis that the mutation rate in this region is
perhaps determined by the three-dimensional chromatin structure than by the
G+C content. (2) our high G+C intergenic region might be embedded in high G+C
genic regions which protect the G+C decay by selection force. In that case, the
relatively low AT-driving-force in the intergenic region is not really neutral.
The limiting isochore conclusion may not be reached if we assume that the mutational
driving force in the current high G+C regions becomes stronger with time,
due to the lower G+C content in the future. However, there is no way to prove
this with the current data.

\section*{Discussion}

\indent

In this study, we re-visit the topic which was popular in the past,
on base composition change \citep{duret,guli,alvarez,romiguier}, but
focus one one species only, the homo sapiens. Towards this, we rely on
the mutational events observed in human only, i.e., the {\sl de novo} mutation
by comparing the genomic sequences between parents and offspring.
A reliable data on both the mutation rate and context-dependence
is clearly important.  Previous work calculated this information
by comparing the orthologous regions between human and chimpanzee,
considering chimp as ancestral and count any single nucleotide change
from chimp to human as mutational events (Supplementary text of  \citep{samocha}).
This approach may obtain more counts, but the directionality
of the mutational events can be questioned. The data we use
is guaranteed for the mutational direction (from parents to offspring) which is
an important piece of information on context analysis.

Rare variants might be another type of data to study mutation
rates and context effect \citep{chakraborty,kimura,neel}.
However, this approach should deal with sequencing errors
and private variant (i.e., variant found only in one person)
 should be validated. Also, population specific
reference genomes should be available, so that a so-called rare
variant according to the standard reference genome might be
not so rare in a particular population, and multiple mutations
on the same site should be corrected. 
Considering the importance of estimating the background neutral
mutational rate in the assessment of excess mutation in a particular
gene \citep{samocha}, it could be interesting to compare different
approaches, with the anticipation of further complexity
due to factors such as gender and age \citep{jonsson},
population \citep{mathieson,durbin}, and chromosome regions \citep{pritchard}.

Our conclusion that isochore-like structure, i.e., different regions
having different G+C contents, can be maintained in the limiting
configuration of the neutral dynamics, has already been implied
in (page 886 of) \citep{lander}: ``if K is the equilibrium constant $\dots$
then the equilibrium GC content should be 1/(1+K) $\dots$ (K) varies
as a function of local GC content". Because the currently (G+C)-rich
regions have lower K (equivalent to our $[\alpha_1(x)/\alpha_2(x)] \cdot [(1-x)/x]$ in
Eq.(\ref{eq-limit}) ), they should also have a higher G+C content in
the limiting equilibrium state. However, our conclusion
is reached based on a more realistic three-variable dynamics
(Eq.\ref{eq-limit3}). We also caution on an assumption required for
reaching this conclusion, i.e. the mutational rate is a chromosome 
regional property and may not be a property of G+C content itself.
To confirm or reject the assumption, it might be necessary to
follow the temporal  base composition dynamics in an intergenic G+C rich region
in the human genome.

\section*{Acknowledgment}

W.L. thanks the financial support from the Robert Boas Center for
Genomics and Human Genetics.

\newpage

\appendix 

\begin{center}
{\bf Appendix}
\end{center}

\section{Derivation of the limiting composition based on mutation rate}

The master equation or continuous time Markov process for the dynamics of
a genomic unit with multiple ($m$)  states is (superscript $T$ is for transpose):
\begin{equation}
\frac{d \vec{ {\bf P} }}{dt} = ({\bf M^T }- {\bf I}) \vec{ {\bf P} }
\label{eq-master}
\end{equation}
where $\vec{ {\bf P}}$ is the composition array with $m$ elements,
and ${\bf M}_{m \times m} = \{ M_{ij} \}=\{ P_{i \rightarrow j} \} (i,j= 1 \dots m)$
is the $m \times m$ transition matrix, with $P_{i \rightarrow j}$ the unit time probability
for state $i$ to change to state $j$, and ${\bf I}$ the $m \times m $ identity matrix.
The value of $m$ is 
4 for nucleotide bases, 16 for dinucleotides, 20 for amino acids,
64 for codons, and any values in between or beyond when degenerate/equivalent states
of the genomic unit are combined. For example, if the strand symmetry is considered, $m=2$;
if A and T are combined into weak and C and G combined to strong, $m=2$;
if C or G within dinucleotide CpG is distinguished from not within, $m=3$;
and if stop codons are excluded from all codons, $m=21$, etc.

Eq.(\ref{eq-master}) can be derived by examining the source of
state-$j$ frequency change at time $t+dt$ from that at time $t$:
\begin{eqnarray}
p_j(t+dt) &=& p_j(t) + \sum_{i \ne j} p_i(t) P_{i \rightarrow j} dt 
 - p_j(t) \sum_{k \ne j} P_{j \rightarrow k} dt \nonumber \\
&=& p_j(t) + \sum_{i \ne j} p_i(t) P_{i \rightarrow j} dt
 - p_j(t) (1-P_{j \rightarrow j}) dt \nonumber \\
&=& p_j(t) + \sum_{i} p_i(t) P_{i \rightarrow j} dt - p_j(t) dt 
\end{eqnarray}
then,
\begin{eqnarray}
\frac{p_j(t+dt)-p_j(t)}{dt} &= & \sum_i p_i(t) P_{i \rightarrow j} - p_j(t)
 \nonumber \\
& =& \sum_{j'} P_{j' \rightarrow i'} p_{j'}(t) - p_j(t) \nonumber \\
& =& 
\sum_{j'}{\bf  M}^T_{i'j} p_{j'}(t) - p_j(t)
\end{eqnarray}
which is Eq.(\ref{eq-master}) in the $dt \rightarrow 0$ limit.

The equilibrium composition is the solution of
$d {\bf \vec{ \bf P}}/dt=0= ({\bf M^T -I}) \vec{ {\bf P}}$ which is
an eigenvalue/eigenvector problem. This type of dynamical systems
is also called (multi) compartmental systems \citep{compart},
and it is known that the only non-negative eigenvalue of
compartmental matrix ${\bf M^T -I}$ is zero (the largest eigenvalue
of matrix ${\bf M^T}$ is one)  \citep{compart}. In other words,
the limiting composition is the (normalized) eigenvector corresponding
to the eigenvalue=1 of the transpose of Markov transition matrix.

\newpage

\begin{table}
\begin{center}
\begin{tabular}{c|c|cccccc|cc|ccc}
\hline
type & n & n(WS) & n(WS$_p$) & n(SW) & n(SS$_p$) & n(S$_p$W) & n(S$_p$S) &  
 $\alpha_1$ & $\alpha_2$ & $\beta_1$ & $\beta_2$ & $\beta_3$ \\
\hline
intron & 58824 & 14100 & 5926 & 18051 & 1108  & 11646  & 523 & 0.597 & 0.403 & 0.434 &  0.326 &  0.241 \\
intergenic& 40375 & 9813 & 4018 & 13277 & 732 & 6705 & 303 & 0.591  & 0.409 &  0.423 &  0.302 &  0.274 \\
missense & 3235 &493&272&875&70&1083&59 & 0.719   & 0.281  &  0.553 &  0.298 &  0.150 \\
stream & 4832 & 1041&513&1505&123&991&44  & 0.616   & 0.384  &  0.449 &  0.294 &  0.257 \\
synonymous & 1336 & 167&112&431&24&508&10 & 0.771  &  0.229  &  0.623 &  0.232 &  0.145 \\
UTR	& 1514 & 359&136&464&34&307&32 & 0.609  &  0.391  &  0.449 &  0.337 &  0.214 \\
non-coding& 634 & 98&68&169&19&171&16 & 0.672  &  0.328  &  0.506 &  0.282 &  0.213 \\
stop	& 149 & 3&1&61&0&66&0 & 0.969  & 0.031  &   0.940 &  0.046 &  0.006 \\
splice	& 81 & 4&16&42&0&1&0  & 0.683  &  0.317  &  0.058 &  0.006 & 0.936 \\
\hline
\end{tabular}
\end{center}
\caption{\label{table1}
Statistics of the {\sl de novo} mutational events used:
type: 9 functional groups are used by simplifying the original 18 groups;
$n$: number of mutational events;
$n(WS)$: number of {\sl de novo} mutation from $W$ (A or T) to $S$ (G or C) bases outside a CpG-containing triplet;
$n(WSp)$: number of {\sl de novo} mutation from $W$ to $S$ bases within a CpG-containing triplet ;
$n(SW)$: number of {\sl de novo} mutation from $S$ (non-CpG)  to $W$ bases;
$n(SSp)$: number of {\sl de novo} mutation from $S$ (non-CpG) to $S$ (CpG containing) bases;
$\alpha_1, \alpha_2$: defined in Eq.(\ref{eq-limit0}), satisfying $\alpha_1+\alpha_2=1$;
$\beta_1, \beta_2, \beta_3$: defined in Eq.(\ref{eq-limit3}), satisfying $\beta_1+\beta_2+\beta_3=1$.}
\end{table}

\begin{table}
\begin{center}
(i) {\sl de novo} mutation in intergenic regions\\
\begin{tabular}{c|c|cccccc|cc|ccc}
\hline
G+C range & n & n(WS) & n(WS$_p$) & n(SW) & n(SS$_p$) & n(S$_p$W) & n(S$_p$S) &
 $\alpha_1$ & $\alpha_2$ & $\beta_1$ & $\beta_2$ & $\beta_3$ \\
\hline
0.1 - 0.35 & 6449 & 2008 & 558 & 2075 & 78  & 698  & 27 & 0.519 & 0.481 & 0.355 & 0.335 &  0.310 \\
0.35- 0.38 & 7288  & 2045 & 724 & 2361  & 109  & 1011  & 31  & 0.549 & 0.451 & 0.382 & 0.320 & 0.299 \\
0.38- 0.40 & 6890 & 1775  & 639 & 2383 & 124 & 905  &  37& 0.577 & 0.423 & 0.401& 0.289& 0.310 \\
0.40 -0.43 & 6125 & 1365 & 620 & 2039 &  114 & 1092  & 48 & 0.612 & 0.388 & 0.443 & 0.287& 0.270 \\
0.43 -0.47 & 6315 & 1285 & 645 & 2193 &  128 & 1260 & 45 & 0.641 & 0.359  & 0.473 & 0.267& 0.260 \\
0.47 - 0.76 & 7308 &  1335&  832 & 2226  &  179 & 1739 & 115 & 0.647 & 0.353 & 0.480 & 0.278 & 0.242 \\
\hline
0.47 - 0.5 & 3136 & 610 & 330 & 985 & 65 & 696 & 38 & 0.641 & 0.359 & 0.476 & 0.285 & 0.239 \\
0.5 - 0.54 & 1964 & 373 & 230 & 602 & 49 & 455 & 28 & 0.637 & 0.363 & 0.469 & 0.279 & 0.252 \\
0.54 - 0.76 & 2208 & 352 & 272 & 639 & 65 & 588 & 49 & 0.663 & 0.337 & 0.496 & 0.265 & 0.239 \\
\hline
\end{tabular}
(ii) equilibrium (G+C)\% andf CpG\% \\
\begin{tabular}{c|cc|ccc}
\hline
G+C range & (G+C)\% & CpG\%/(G+C)\% & eq(G+C) by Eq.(\ref{eq-limit}) & eq(G+C) by Eq.(\ref{eq-limit3}) & eq CpG/(G+C) by Eq.(\ref{eq-limit3})\\
\hline
0.1 - 0.35 & 0.343 & 0.0309 & 0.326 & 0.329 & 0.0286  \\
0.35- 0.38 & 0.365 & 0.0320 & 0.321 & 0.325 & 0.0300 \\
0.38- 0.40 & 0.389 & 0.0355& 0.319 & 0.315 & 0.0380 \\
0.40- 0.43 & 0.413 & 0.0418 & 0.309 & 0.313 & 0.0394 \\
0.43- 0.47 & 0.447 & 0.0518 & 0.311 & 0.313 &  0.0506 \\
0.47- 0.76 & 0.496 &  0.0667 & 0.350 & 0.361 &  0.0586\\
\hline
0.47 - 0.5 & 0.482 & 0.0618 & 0.343 & 0.356 & 0.0524 \\
0.5 - 0.54 & 0.513 & 0.0721 & 0.376 & 0.384 & 0.0653 \\
0.54 - 0.76 & 0.560 & 0.0932 & 0.393 & 0.403 & 0.0849 \\
\hline
\end{tabular}
\end{center}
\caption{\label{table2}
Statistics of intergenic {\sl de novo} mutations partitioned into their 2kb window G+C content range (6 bins
for rough equal number partition, and last bin further split into three bins):
(i) see the caption of Table \ref{table1};
(ii) predicted limiting equilibrium by current G+C content using  Eq.(\ref{eq-limit}),
or by both current G+C and current CpG/(G+C) using  Eq.(\ref{eq-limit3}).}
\end{table}

\end{document}